# Multiphonon interactions and collective excitations in charge density wave of transition metal dichalcogenides

Divya Rawat, Juhi Pandey, and Ajay Soni[*]

*School of Physical Sciences, Indian Institute of Technology Mandi, Mandi, 175005, HP India*

*Email:* ajay@iitmandi.ac.in

Charge density wave (CDW) is a periodic modulation of electronic charge leading to a reconstruction of the lattice, an emergence of zone folded mode along with collective excitations.[1] Transition metal chalcogenides have shown a great potential to study the underlying physics of multi body interactions like electron-phonon (*e-ph*) as well as plasmons-phonon coupling,[2,3] excitonic complexes[4] and CDW instabilities.[5,6] We are elaborating on the collective excitation and multiphonon interactions like amplitude, zone-folded and two phonon ($2_{ph}$) mode associated with CDW in 2H-TaS$_2$ and 2H-NbSe$_2$. The $2_{ph}$ is a unique excitation appears due to anharmonicity mediated longitudinal acoustic modes associated with the Kohn anomaly and provides a substantial evidences of *e-ph* coupling.[7] The experimental observations of anomalous response of $2_{ph}$ mode across CDW is clarified with phonon self-energy formalism. Transport measurements shows that both materials exhibit incommensurate CDW ($T_{I-CDW}$) at ~ 76 K (2H-TaS$_2$) and ~ 30 K (2H-NbSe$_2$). As revealed by the estimated *e-ph* coupling ($\lambda$ ~ 0.007 for 2H-NbSe$_2$ and ~ 0.013 for 2H-TaS$_2$) and anharmonicity constant ($\delta$ ~ 6.92 for 2H-NbSe$_2$ and ~ 2.723 for 2H-TaS$_2$), we emphasize on higher *e-ph* coupling in 2H-TaS$_2$ while larger anharmonicity 2H-NbSe$_2$.

Two phonon (*2$_{ph}$*) modes are the collective excitations associated with the correlation of multiple phonon branches in the materials having charge density wave (CDW) instabilities arising due to the reconstruction of lattice. CDW is a quantum phenomenon, proposed by Peierls in 1955, wherein the periodic modulation of the electronic charge density is accompanied by the lattice distortion and high anharmonicity.[1] Based on the periodicity of the distorted lattice, CDW can be classified as commensurate (C-CDW), and incommensurate (I-CDW) CDW, where the ordering vector ($q_{CDW}$) is an integral and non-integral multiple of the reciprocal lattice vector of the undistorted phase, respectively.[8] In general, the origin of CDW can be explained by Fermi surface nesting (FSN),[1,5] electron-phonon (*e-ph*) coupling,[9] anharmonicity,[7] and saddle-point singularity.[10] In undistorted solid, the energy spectrum of the electrons consists of allowed energy levels separated by an energy band gap at Brillouin zone (BZ) boundary, (Fig.1a), while in the distorted CDW state the lattice modulation folds the BZ with opening of a gap ($\Delta_{CDW}$) at the Fermi level, ($\vec{q} = \pm\overrightarrow{2k_F}$) (Fig.1b).[11] In the zone-folded (ZF) spectrum, the CDW instability executes new quasi-harmonic vibrations forming a condensate of phonons of specific wave vector, $\pm q$, which are named as collective (amplitude/phase) modes.[12] One such collective modes is formed when an incident photon excites an electron-hole pair which couples to an optical phonon through *e-ph* coupling and named as $2_{ph}$ mode.[11] The Feynman diagram representing the Raman scattering of $2_{ph}$ mode is shown in Fig.1c.[7] Specifically, the $2_{ph}$ mode is a second order scattering near $\vec{q}_{CDW} = 2/3\Gamma M$ of two acoustic mode with equal and opposite momentum. The scattering lowers the energy of the modes at certain symmetry points of BZ and appears as a discontinuity (called as Kohn anomaly) in the slope of phonon dispersion (Fig.1d), with a frequency approximately half of $2_{ph}$ mode.[11] The $2_{ph}$ mode has been observed in graphene,[13] group-IV nitrides[14] and group-V carbides[15] using Raman scattering and also supported by density functional theory. For transition metal dichalcogenides (TMDCs), the Kohn anomaly arise due to the presence of the coupling of *d* electrons of transition metal with phonons.

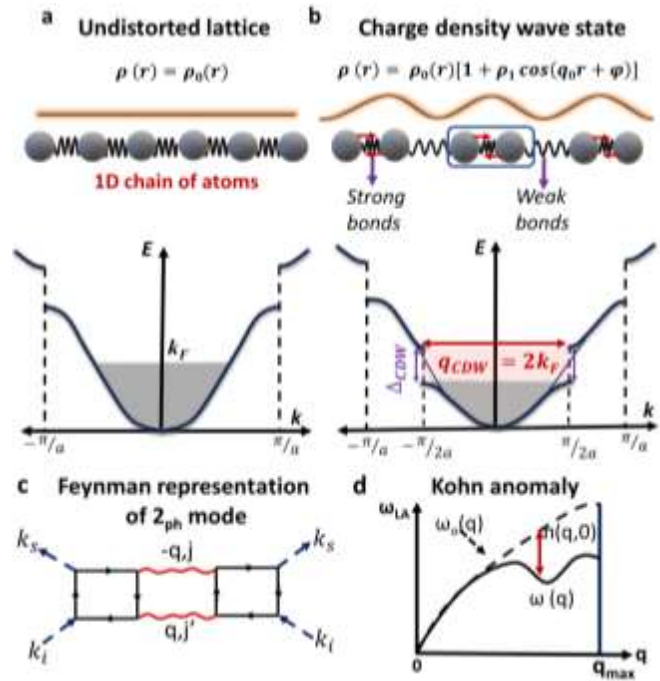

**Fig. 1 | Schematic of charge density wave and two-phonon mode. a-b,** The atomic arrangement, electronic charge density, and energy band diagram of **a,** undistorted lattice, and **b,** charge density wave state. Modulation of electronic density with change in potential energy of lattice resulting in the opening of band gap ($\Delta_{CDW}$) at fermi level, $q = 2k_F$. **c,** Feynman representation for the Raman scattering of $2_{ph}$ mode, **d,** The anomalous dispersion curve for longitudinal acoustic (Kohn anomaly) mode with inclusion of all interaction in terms of phonon self-energy ($\pi(q,0)$ from the *d* electrons of metal atoms.

For TMDCs, while both FSN[1,5] and *e-ph* coupling[9] can drive the CDW yet various experiments like electrical and thermal transport measurement,[16] magnetic measurement,[17] and neutron scattering,[18] and Raman spectroscopy,[9] advocate the *e-ph* coupling as the major cause of CDW instabilities. Previous experimental and theoretical studies such as inelastic neutron and X-ray scattering[18,19], time resolved optical pump-probe spectroscopy[20], density functional calculation[19], on TMDCs showed the softening of the acoustic and $2_{ph}$ mode with lowering of temperatures until CDW, but the behavior of $2_{ph}$ mode across CDW is unclear. In this regards, the interpretation of CDW with collective and $2_{ph}$ mode, especially in case of 2H-NbSe$_2$, 2H-TaSe$_2$,[21] 2H-TaS$_2$,[22] 1T-VSe$_2$[8] is little explored so far.[23] Thus, we report on the involvement of $2_{ph}$ mode with *e-ph* coupling and unveiling the key fundamental role of $2_{ph}$ mode on the onset of CDW phase transitions in 2H-TaS$_2$ and 2H-NbSe$_2$, using low temperature and polarized Raman spectroscopy. We emphasized on the multiphonon interactions associated with CDW instabilities using light matter interactions, which are a powerful way to identify the nature of collective excitations, ZF modes as well as $2_{ph}$ mode.[9,24] The growth and characterization of single crystal 2H-TaS$_2$ and 2H-NbSe$_2$ are elaborated in the methods, below.

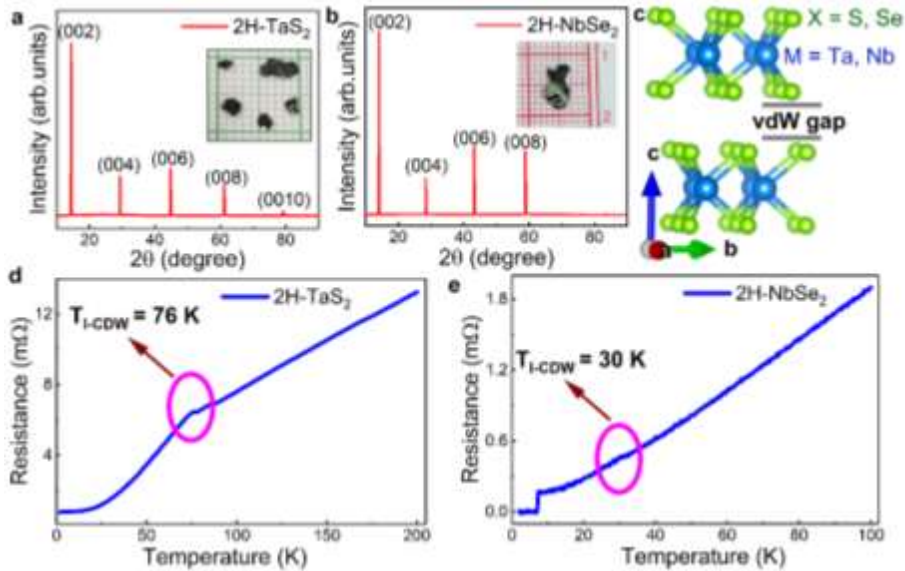

**Fig. 2| Characterization of single crystal and observation of CDW.** X-ray diffraction pattern of single-crystals **a,** 2H-TaS$_2$ and **b,** 2H-NbSe$_2$, (inset show the photograph of the grown crystals) and **c,** crystal structure of the 2H-MX$_2$ (M = Ta, Nb; X = S, Se), temperature dependent resistance (*R(T)*) of **d,** 2H-TaS$_2$ ($T_{I\text{-}CDW}$ at ~ 76K) and **e,** 2H-NbSe$_2$ ($T_{I\text{-}CDW}$ ~ 30 K).

Indian Institute of Technology Mandi

The single crystalline nature of 2H-TaS$_2$ and 2H-NbSe$_2$ (both have space group *P6$_3$/mmc*, 194) has been confirmed from X-ray diffraction pattern (Fig. 2a-b), and the miller indices along (*00l*) direction of all peaks indicate the growth along the *c*-axis. The estimated lattice parameter are $a = b = 3.316$ Å and $c = 12.568$ Å (for 2H-TaS$_2$) and $a = b = 3.442$ Å and $c = 12.548$ Å (for 2H-NbSe$_2$), which are in agreement with the reports.[9,25] Figure. 2c shows the hexagonal unit cell of 2H-MX$_2$ with covalently bonded tri-atomic layers (*X-M-X*), where metal (*M = Ta* and *Nb*) atoms are sandwiched between two layers of chalcogens (*X = S* and *Se*),[26] and the tri-atomic layers are stacked with weak van der Waals gap. Both 2H-TaS$_2$ and 2H-NbSe$_2$ show metallic behavior in the temperature dependent four-probe resistance (*R(T)*) shown in Fig. 2d-e and these materials undergo an I-CDW and the transition temperature (*T-$_{ICDW}$*) can be identified from the change in slope of electrical transport measurements.[27-29] Thus, characteristic $T_{I\text{-}CDW}$ has been observed at ~ 76 K (for 2H-TaS$_2$) and ~ 30 K (for 2H-NbSe$_2$), whereas 2H-NbSe$_2$ also has a superconducting transition, at T$_{SC}$ ~ 7 K, as shown in Fig. 2d-e.

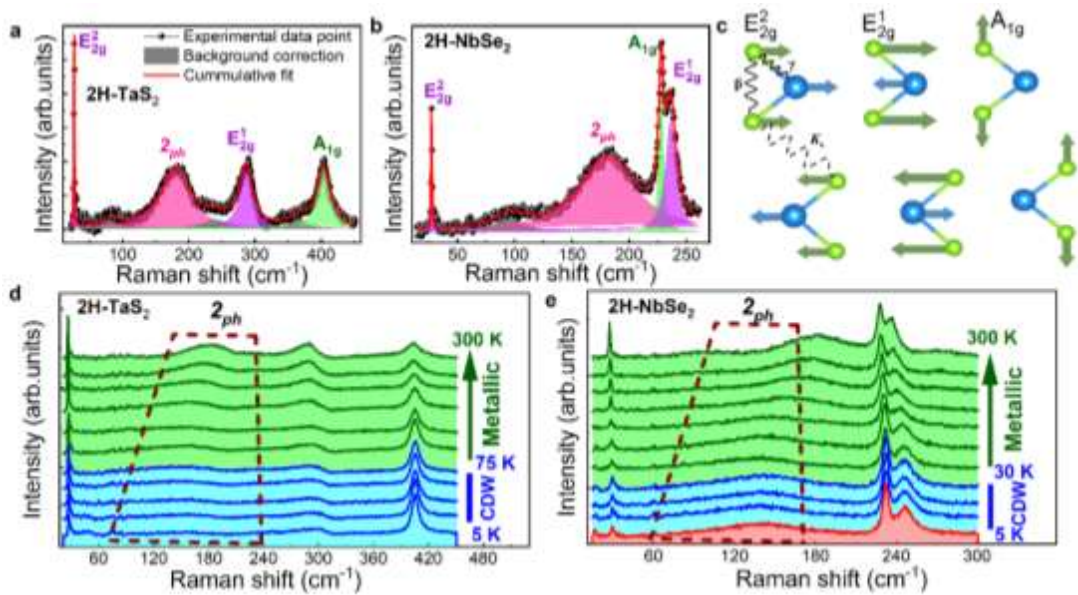

**Fig. 3| Raman spectra of 2H-TaS$_2$ and 2H-NbSe$_2$ single crystals.** Room temperature Raman spectra of **a,** 2H-TaS$_2$ and **b,** 2H-NbSe$_2,$ respectively, and **c,** Schematic of atomic displacements for $E_{2g}^2$, $E_{2g}^1$, and $A_{1g}$, Raman modes. Solid black and dashed curved lines represent the intralayer and interlayer spring constant, respectively. Temperature dependent Raman spectra of **d,** 2H-TaS$_2$ and **e,** 2H-NbSe$_2$.

The primitive cell of 2H-MX$_2$ has three atoms, which correspond to nine vibrational modes at the center of the BZ, three of which are acoustic and six are optical. From group theoretical calculations, optical modes can be represented as, $\chi = A_{1g} + E_{1g} + 2E_{2g} + A_{2u} + E_{1u}$, where $A_{1g}$, $E_{1g}$, and $E_{2g}$ are Raman active and $A_{2u}$, and $E_{1u}$ are infra-red active modes[30]. The room temperature Raman spectra for 2H-TaS$_2$ and 2H-NbSe$_2$ are shown in Fig. 3a-b. Here, 2H-TaS$_2$ depicts inter-layer breathing mode ($E_{2g}^2$) ~ 26 cm$^{-1}$, $2_{ph}$ mode ~ 182 cm$^{-1}$, $E_{2g}^1$ ~ 290 cm$^{-1}$, and $A_{1g}$ ~ 404 cm$^{-1}$ while 2H-NbSe$_2$ have $E_{2g}^2$ ~ 28 cm$^{-1}$, $2_{ph}$ mode ~ 184 cm$^{-1}$, $A_{1g}$ ~ 227 cm$^1$, and $E_{2g}^1$ ~ 236 cm$^{-1}$.[31] The $A_{1g}$ mode involves the out-of-plane vibration (along *c*-axis) of *S*/*Se* atoms against each other whereas $E_{2g}$ mode have in-plane vibrations of *Ta* (*Nb*) and *S* (*Se*) atoms, and the schematics are presented in Fig. 3c. The other possibilities of Raman active modes have been verified with 633 and 785 nm laser excitations using polarized Raman spectroscopy and are presented in supplementary information (SI) (Fig. S1). The energies of $A_{1g}$ and $E_{2g}$ modes for 2H-TaS$_2$, are higher than 2H-NbSe$_2$ due to stronger ionic bonding arising from higher electronegativity of *S* atom compared to *Se*.[32] Besides the characteristic $A_{1g}$ and $E_{2g}$ modes, a broad CDW signature, $2_{ph}$ mode has also been observed for both 2H-TaS$_2$ and 2H-NbSe$_2$. The detailed analysis of the $2_{ph}$ mode is done in later texts.

Figure. 3d-e shows the low-temperature Raman spectra for 2H-TaS$_2$ and 2H-NbSe$_2$, respectively. For 2H-NbSe$_2$, the temperature variation of phonon frequency ($\omega(T)$) of both $A_{1g}$ ~ 227 cm$^1$, and $E_{2g}^1$ ~ 236 cm$^{-1}$ mode exhibiting hardening because of thermal expansion and anharmonicity (Fig. S2). For 2H-TaS$_2$, the $\omega(T)$ of $A_{1g}$ ~ 404 cm$^{-1}$ mode illustrate anharmonic behavior, while the $E_{2g}^2$ ~ 290 cm$^{-1}$ mode show peculiar response (Fig. S2), where negligible shift has been observed down to ~ 100 K, and then hardens with lowering of T, below 100 K.[9] Such an anomalous response of the $E_{2g}^2$ mode indicates the signatures of *e-ph* coupling, associated with the lattice distortion in the CDW phase along the in-plane direction.[9] Further, in line with literatures, various collective modes due to CDW superlattices have also been observed below CDW transition (Fig. S3). In bulk 2H-TaS$_2$, CDW amplitude mode ~ 48 cm$^{-1}$ ($E_{2g}^2$), ~ 77 cm$^{-1}$ ($A_{1g}$), and ZF mode ~ 95 cm$^{-1}$ are observed (Fig. S3a), while 2H-NbSe$_2$ has amplitude mode ~ 38 cm$^{-1}$, ZF mode ~ 92 cm$^{-1}$, and a weak mode ~ 190 cm$^{-1}$, (Fig. S3b) and the origin is unknown so far. The ZF modes of 2H-TaS$_2$ ~ 95 cm$^{-1}$ and 2H-NbSe$_2$ ~ 92 cm$^{-1}$ are positioned at approximately half of the frequency of their respective $2_{ph}$ mode, and hence, their origin is associated with the Kohn anomaly.

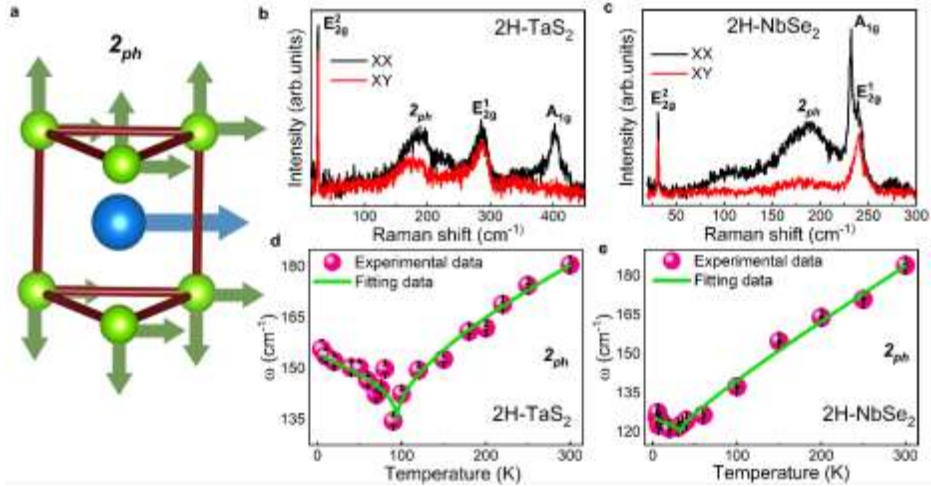

**Fig. 4| Involvement of two-phonon mode ($2_{ph}$) in charge density wave. a,** Atomic displacement of $\Sigma_1$ symmetry associated with $2_{ph}$ mode. Polarized Raman spectra of **b,** 2H-TaS$_2$, and **c,** 2H-NbSe$_2$, (*XX* is parallel and *XY* is perpendicular polarization). The anomalous variation of the $2_{ph}$ mode for **d,** 2H-TaS$_2$ and **e,** 2H-NbSe$_2$, indicates the softening with decreasing temperature until T ~ 100 K and T ~ 30 K, respectively, and subsequent hardening at low temperatures.

The $2_{ph}$ mode represents a second-order scattering of longitudinal acoustic modes from the other side of the BZ thus exhibits a very low Raman intensity.[33] However, the existence of singularity at $q \sim 2k_F$ in the electronic polarizability can enhance the intensity comparable to characteristic modes. The $2_{ph}$ mode in 2H-TaSe$_2$,[34] 2H-NbSe$_2$,[34] and 2H-NbS$_2$[35] have a unique $\Sigma_1$ symmetry, which predominantly arises because of longitudinal vibration of chalcogen and metal atoms.[18] The schematic of $\Sigma_1$ symmetry of $2_{ph}$ mode is presented in Fig. 4a, where the in-plane displacements of metal atom are larger than chalcogen atom.[34] The involvement of both longitudinal and transverse displacements of $2_{ph}$ mode have been further confirmed by the polarized Raman spectroscopy using 532 nm laser, at 300K (Fig. 4b-c), where $A_{1g}$ mode disappeared in *XY* configuration, while $2_{ph}$ mode with reduced intensity is discernible. The excitation energy dependence of the longitudinal vibrations of $2_{ph}$ mode have also been confirmed in polarized Raman spectra (*XY* configuration) using 633 and 785 nm lasers (Fig. S1). Therefore, we believe that $2_{ph}$ mode involve predominant longitudinal vibration of atoms and have a strong influence on the CDW superlattice forms in the basal plane of the hexagonal BZ.[7]

The $2_{ph}$ mode shows anomalous softening with cooling before the onset of $T_{I-CDW}$ for 2H-NbSe$_2$ and hardens later (Fig. 4d-e), on the other hand for 2H-TaS$_2$, the CDW stabilities can exists even at higher temperatures,[9] which is realized from diminished softening at ~ 100 K. The anomalous softening and hardening of $2_{ph}$ mode below $T_{I\text{-}CDW}$ arising due to the anharmonicity along with interaction of phonon with the *d* electron of *Ta* and *Nb*.[34] We have fitted the $\omega(T)$ of the $2_{ph}$ mode for both 2H-TaS$_2$ (Fig. 4d) and 2H-NbSe$_2$ (Fig. 4e), by mean field theory considering a combination of (i) Klemens model for normal mode and (ii) second-order soft mode theory by Tsang.[36] With inclusion of momentum conservations, the combination can be described as $\Delta\omega(T) = \omega_o - \delta\left(\frac{2}{e^{\frac{\hbar\omega o}{2k_BT}}-1}\right) + \lambda\omega_0'|T - T_{cf}|^{\frac{1}{2}}$, where $\omega_0$ (at ~ 0K) and $\omega_0'$ (at high temperatures) of $2_{ph}$ mode, $\delta$ is an anharmonic constant (depends on phonon dispersion), $\lambda$ is *e-ph* coupling constant and $T_{cf}$ is the transition temperature of instabilties.[36] The experimental data are well fitted with equation and are showing a significant softening down to ~ 100 K (2H-TaS$_2$) and ~ 30 K (2H-NbSe$_2$), respectively. Here, the value of $\omega_0'$ is fixed and the extracted parameters are tabulated in Table S1. of SI. For 2H-NbSe$_2$, the $T_{cf}$ ~ 32 K is approximately equivalent to $T_{I\text{-}CDW}$ ~ 30 K, whereas $T_{cf}$ of 2H-TaS$_2$ is ~ 100 K, which is higher than the $T_{I\text{-}CDW}$ ~ 76 K showing the involvements of instabilities. The high $T_{cf}$ indicates that CDW instability in 2H-TaS$_2$ can persists above the actual $T_{I\text{-}CDW}$ (~ 76 K), which is also established by the anomalous $\omega(T)$ of $E_{2g}^2$ mode (Fig. S2). The observation of high $T_{cf}$ $2_{ph}$ mode for 2H-TaS$_2$ has also been shown by the time-resolved optical pump-probe spectroscopy and density functional theory, which predict the existence of non-BCS like transition at ~ 100 K.[7,20] The bulk 2H-TaS$_2$ have $\delta$ ~ 2.723 and $\lambda$ ~ 0.013, whereas 2H-NbSe$_2$ have $\delta$ ~ 6.92 and $\lambda$ ~ 0.007, which is indicating that both materials have significant amount of anharmonic interactions. On comparing the value of $\lambda$, it has been observed that strength of *e-ph* coupling is stronger in case of 2H-TaS$_2$ than 2H-NbSe$_2$ leading to persistence of CDW instability in 2H-TaS$_2$ at higher temperature (~ 100 K). Unlike to the earlier reports on inelastic X-ray scattering of 2H-NbSe$_2$[19] and VSe$_2$[37], where the longitudinal acoustic mode have been collapsed in the CDW region whereas we have observed an anomalous hardening for $2_{ph}$ mode of 2H-TaS$_2$ and 2H-NbSe$_2$. Our study illustrates phonon hardening below $T_{I\text{-}CDW}$, which can be incorporated with the numerical formalism of phonon self-energy in the presence of anharmonic interaction and *e-ph* coupling.[7,11] (Section.4 in SI) The anomalous $\omega$ (*T*) and full-width at half-maximum ($\beta$) of $2_{ph}$ mode, (Fig. S4a) increase till ~ 100 K and then decreases. Such an anomalous behavior of $2_{ph}$ mode arises from the coupling of phonons with *d* electrons near the fermi energy and due to the anharmonic interaction.[11,19] Thus, not only the collective modes the anharmonic mediated $2_{ph}$ mode plays a key role in understanding of CDW superlattices of 2H-polytypes of 2H-TaS$_2$ and 2H-NbSe$_2$.

In summary, the origin, symmetry, and key role of multiphonon interactions (collective amplitude and anharmonic-mediated $2_{ph}$ mode) across the CDW instability are emphasized for 2H-TaS$_2$ and 2H-NbSe$_2$. The low temperature polarized Raman results provide considerable evidence for the existence of CDW instability in 2H-TaS$_2$ even up to 100 K, which is high as compared to the $T_{I\text{-}CDW}$ ~ 76 K. The observations clarified the involvement of longitudinal vibration of atoms associated with $2_{ph}$ mode and key evidence of *e-ph* coupling with anharmonicity during reconstruction of the CDW lattices.

## Methods

Single crystal of 2H-NbSe$_2$ and 2H-TaS$_2$ were grown by chemical vapor transport technique using iodine as a transporting agent.[38] X-ray diffraction of the grown single crystal was obtained using a rotating anode Rigaku SmartLab diffractometer in Bragg-Brentano with CuK$_α$ radiation (λ = 1.5406 Å). Low temperature resistance measurement was performed by four probe method in the temperature range of 3 to 300 K using a Quantum Design make physical properties measurement system. Raman scattering measurements were performed using Horiba Jobin-Vyon LabRAM HR Evolution Raman spectrometer with Czerny-turner grating (1800 gr/mm) and Peltier cooled CCD detector in a back-scattering configuration. The 532 nm, 633 nm, and 785 nm lasers were used to excite the sample and 50X long working distance objective lens was used to focus on the crystal. Ultra-low frequency filters were used to access low-frequency Raman modes. To control the polarization, a λ/2 half-waveplate and an analyzer have been used before the objective lens and spectrometer respectively to select the desired polarization of the incident and scattered light. Temperature-dependent Raman measurements were performed in the temperature range of 3-300K using a Montana instrument make closed-cycle cryostat. All the Raman spectra were fitted by Lorentzian function to evaluate full width at half-maximum (*β*), frequency (*ω*), and intensity (*I*) of Raman modes.

**Data availability:** All other data supporting the findings of this study are available upon request from the corresponding author.

**Acknowledgment:** We would like to acknowledge IIT Mandi for the research facilities.

**Supplementary Information:** Supplementary Figures. S1-S4 and Sections 1-6.

# Supplementary Information

## Multiphonon interactions and collective excitations in charge density wave of transition metal dichalcogenides

Divya Rawat, Juhi Pandey, and Ajay Soni[*]

*School of Physical Sciences, Indian Institute of Technology Mandi, Mandi, 175005, HP India*

*Email:* ajay@iitmandi.ac.in

This supporting file has the experimental data and explanation which is complimentary to the main manuscript.

**1. Polarized Raman spectra of 2H-TaS$_2$ and 2H-NbSe$_2$ using 633 nm and 785 nm laser.**

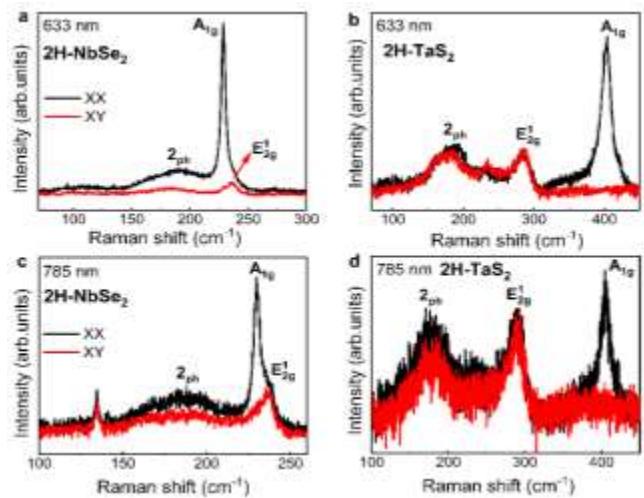

**Figure S1: Polarized spectrum with different laser excitation.** Polarized Raman spectra for 2H-NbSe$_2$ and 2H-TaS$_2$ with **a-b**, 633 nm, and **c-d**, 785 nm laser, respectively.

Polarized Raman spectroscopy with different laser 633 nm, and 785 nm have been used to observe various Raman modes in both parallel (*XX*) as well as perpendicular polarization (*XY*).[1] For 2H-NbSe$_2$, all characteristics modes like $A_{1g}$, $2_{ph}$ and $E_{2g}^1$ are observed with both 532 nm (Fig. 3d), and 633 nm (Figure. S1a), whereas an extra mode at ~ 135 cm$^{-1}$ also been observed only with 785 nm (Figure. S1c). This ~ 135 cm$^{-1}$ mode has an *E*-symmetry as it is present in both *XX* and *XY* configuration. Similarly, for 2H-TaS$_2$ (Figure. S1b and d), all characteristic modes are intense while the $E_{1g}$ ~ 235 cm$^{-1}$ mode is visible only in *XY* configuration.[2] The mode ~ 235 cm[-1] is predicted by Zhang *et.al* but not observed so far clearly.[2,3]

2. **Variation of phonon frequency ($\omega$) and full-width at half-maximum of $A_{1g}$ and $E_{2g}^1$ mode of 2H-TaS$_2$ and 2H-NbSe$_2$.**

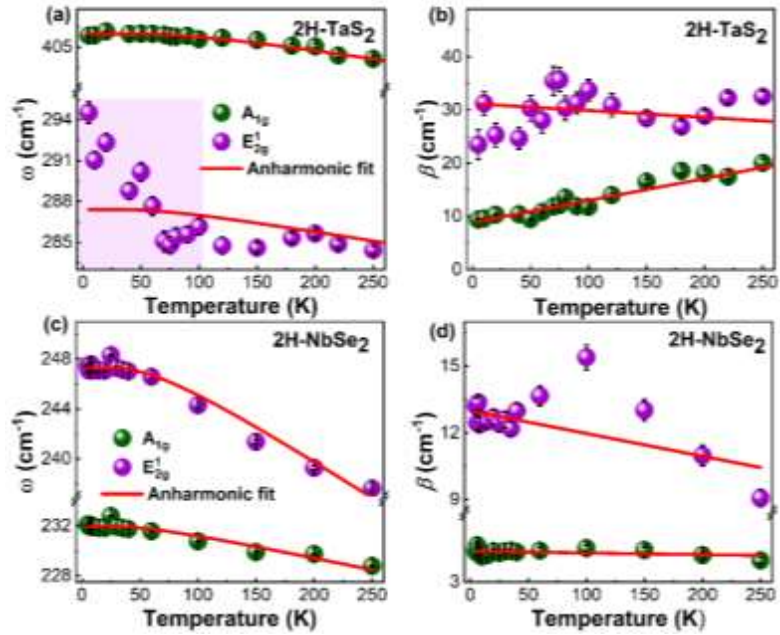

**Figure S2: Temperature dependence of characteristic modes.** Temperature dependent $\omega$ and *full-width at half-maximum* of $A_{1g}$ and $E_{2g}^1$ mode for **a-b,** 2H-TaS$_2$ and **c-d,** 2H-NbSe$_2$, respectively.

The $\omega$ and the full-width at half-maximum ($\beta$) of both $A_{1g}$ and $E_{2g}^1$ mode (Figure. S2a-b) for 2H-NbSe$_2$ (Figure. S2c-d) shows the usual anharmonic behavior arising because of thermal fluctuation and anharmonic potential. The two parameters ($\omega$ *and* $\beta$) can be modeled using equation: $\omega = \omega_0 - \left(\frac{A}{e^x - 1}\right)$, *and* $\beta = \beta_0 + \left(\frac{A}{e^x - 1}\right)$; $x = \hbar\omega_o/2k_BT$), where the first term corresponds to the $\omega_0$ ($\beta_0$) at T = 0 K and second term represents multi-phonon processes.[1] For, 2H-NbSe$_2$, the variation of $\omega$ *for* both $A_{1g}$ and $E_{2g}^1$ mode is well fitted with the model, however $\beta$ of $E_{2g}^1$ deviates. In case of 2H-TaS$_2$, only $A_{1g}$ mode fitted well while $E_{2g}^1$ have significant deviation below 100 K. These results indicate that in-plane mode is more sensitive to the formation of CDW superlattice.

3. **Observation of CDW collective modes using 532 nm, 633 nm and 785 nm laser.**

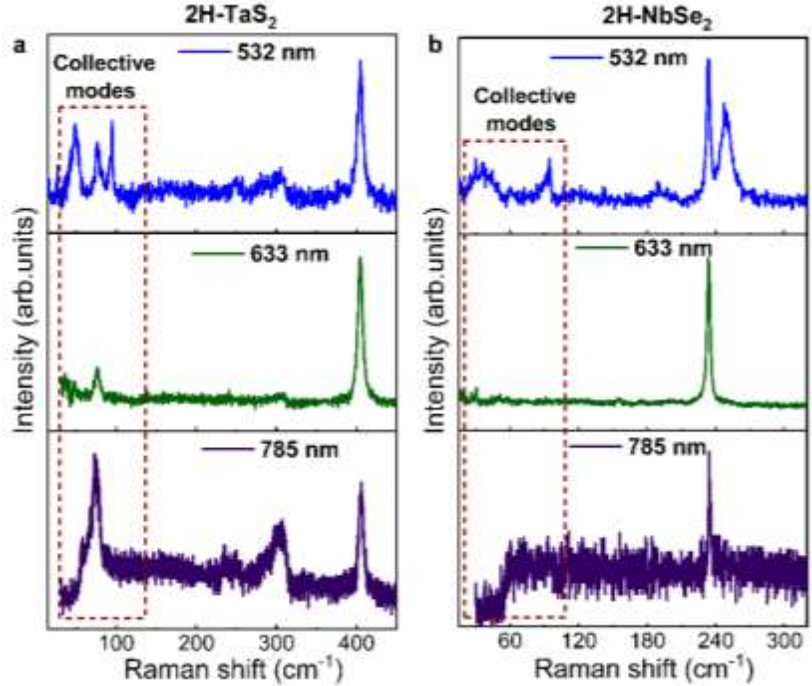

**Figure S3: Observation of collective modes.** Low temperature Raman spectra of **a,** 2H-TaS$_2$ and ( **b,** 2H-NbSe$_2$ with 532 nm, 633 nm, 633nm and 785 nm laser at 3 K.

Bulk 2H-TaS$_2$ exhibits two collective modes, ~ 48 cm$^{-1}$, and ~ 77 cm$^{-1}$, and higher intense zone-folded mode ~ 95 cm$^{-1}$.[3] on the other hand, the bulk 2H-NbSe$_2$ exhibit collective modes at ~ 38 cm$^{-1}$, zone-folded modes 92 cm$^{-1}$, and another weak mode ~ 190 cm$^{-1}$. Collective modes (amplitude and phase mode) mainly originate from the formation of CDW superlattice.

4. **Numerical analysis of the response of two-phonon ($2_{ph}$) mode in both normal and CDW phase.**

The CDW state is accompanied by a distortion of the atomic arrangements in lattice which indicates that these systems are highly anharmonic. The numerical problem of CDW system is formulated by Behera and Mohanti[4] and the Hamiltonian is given by:

$$H = H_{electron} + H_{phonon} + H_{e\text{-}ph} + H_{anharmonic}$$

$$[H = \sum_{k\sigma} \varepsilon_k C^\dagger_{k\sigma} C_{k\sigma} + \sum_q \omega_q b^\dagger_q b_q + g \sum_{qk\sigma} C^\dagger_{k+q\sigma} C_{k\sigma} A_q + \alpha \sum_{q_1 q_2 q_3} A_{q_1} A_{q_2} A_{q_3}].$$

where $C_{k\sigma}^\dagger C_{k\sigma}$ describe the annihilation and creation operator for an electron with momentum $k$, spin $\sigma$ and energy $\varepsilon_k$. The $b_q^\dagger b_q$ describe the corresponding annihilation and creation operator for phonons with wavevector $q$ and frequency $\omega_q$. The $g$ is the electron-phonon coupling constant, $\alpha$ is the anharmonic strength, and $A_q = b_q + b_q^\dagger$. To incorporate the temperature dependence of $2_{ph}$ modes above the CDW transition, the $2_{ph}$ Green's function has to be calculated in presence of *e-ph* interaction using Hamiltonian without the anharmonic term. With these assumptions, the phonon self-energy given by[4]

$$\Sigma_0(\omega) = 36\pi\alpha^2\omega_0\beta_0(\omega) = 432\alpha^2\omega_0 \coth\left(\frac{\beta\omega_Q}{2}\right)\left\{\frac{2\omega_Q}{\omega^2-4\Omega_Q^2(T)} - \frac{1}{8\omega_D}\left[1+\frac{1}{4}\left(\frac{\omega}{\omega_D}\right)^2 \ln\left(1-4\left(\frac{\omega}{\omega_D}\right)^{-2}\right)\right]\right\},$$

where, $\omega_D$ is the Debye frequency, $\omega_0$ is the frequency of the $q = 0$, $\omega_Q$ is the frequency of the original acoustic phonon at q = Q and $\Omega_Q$ is the usual Kohn anomaly phonon frequency which is given by: $\Omega_Q^2(T) = s\omega_Q^2 \ln(T/T_p)$ with the dimensionless coupling constant; $s = 4g^2N(0)/\omega_Q$ and the Peierls transition temperature: $T_p = 1.14E_B k_B^{-1} \exp(-1/s)$. Here, it is clear that phonon self-energy has a pole around $2\Omega_Q$, which appear as a $2_{ph}$ mode in the phonon spectral function with frequency doeble of the Kohn anomaly mode and show the remarkable softening with lowering the temperature. This anharmonic mediated $2_{ph}$ mode would be similar to the one observed in the Raman scattering experiment.

On the other hand during the CDW state, periodic modulation of the electronic charge density is accompanied by the lattice distortion which opens up a CDW gap in the electronic energy spectrum at the Fermi level. To incorporate the behavior of $2_{ph}$ mode in CDW phase, anharmonic interaction are also included. Hence, $H_{CDW}$ in the CDW state is given by:

$$H_{CDW} = \sum_{k\sigma}\varepsilon_k C_{k\sigma}^\dagger C_{k\sigma} - G_0 \sum_{k\sigma} C_{k+Q\sigma}^\dagger C_{k\sigma} - G_1 \sum_{k\sigma}^n C_{k+Q\sigma}^\dagger C_{k\sigma} + G_0 G_1/\lambda,$$

where $G_0$ and $G_1$ are related to the energy gap in the electron spectrum, $\lambda$ is the strength of the electron-electron interaction mediated by the phonons. The assumptions have been taken because of the nesting property $k \pm 2Q = k$ and condition for Raman scattering, which required q = 0. But in the CDW state, the wave vector Q is a reciprocal lattice vector (when zone folded) coincides with the zone-center (q = 0). With these assumptions, the phonon self-energy in the CDW state at zero temperature is given by:[4]

$$\Sigma_Q(\omega) = 4\omega_Q \left\{\frac{3\alpha G}{g} + g^2 N(0)\left[\frac{(4G^2-\omega^2)^{1/2}}{\omega}tan^{-1}\frac{\omega}{(4G^2-\omega^2)^{1/2}} - \ln\left(\frac{2E_B}{G}\right)\right] + 216\alpha^2\left[\frac{\omega_Q}{\omega^2-4\Omega_Q^2} - \frac{1}{16\omega_D}\left\{1+\frac{1}{4}\left(\frac{\omega}{\omega_D}\right)^2 \ln\left(1-4\left(\frac{\omega}{\omega_D}\right)^{-2}\right)\right\}\right]\right\}.$$

Here, there is a square-root singularity at ω = 2G and $2\Omega_Q$, which will develop into the CDW amplitude and two-phonon mode.[5] Here, the two-phonon mode exhibit hardening decreasing temperature (below CDW), which is supporting our experimental data.

## 5. Variation of β and Raman intensity of $2_{ph}$ mode of 2H-TaS$_2$ and 2H-NbSe$_2$.

**Table S1: Extracted parameter from the ω(T) (Fig. 4(d-e)) for both 2H-TaS$_2$ and 2H-NbSe$_2$**

|  | $\omega_0$ | $|\gamma|$ | $|\lambda|$ | $\omega_0'$ | $T_c$ |
|---|---|---|---|---|---|
| **2H-TaS$_2$** | 128.76 ± 1.59 | 2.723 ± 0.85 | 0.013 ± 0.0023 | 182 | 94.18 ± 3.48 |
| **2H-NbSe$_2$** | 111.38 ± 1.14 | 6.92 ± 1.04 | 0.007 ± 0.0025 | 184 | 30.35 ± 3.49 |

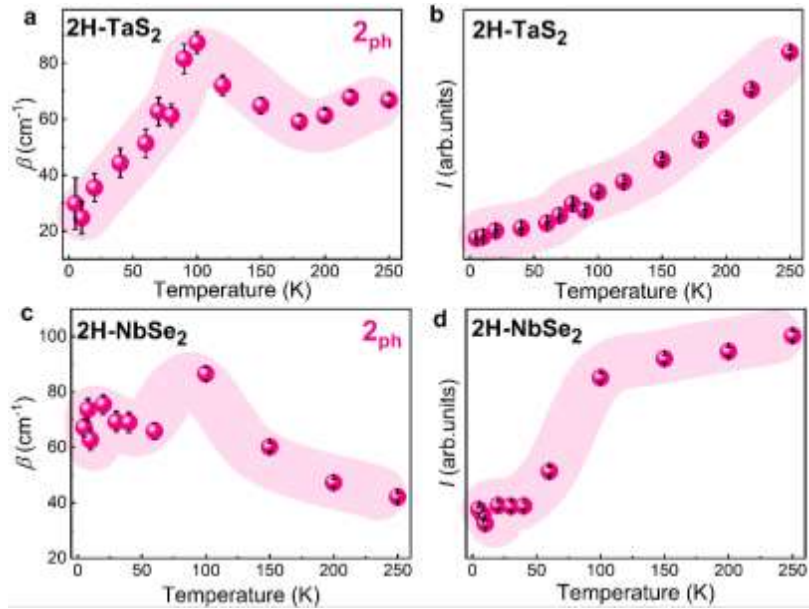

**Figure S4: Temperature dependent response of $2_{ph}$ mode.** Temperature dependent β and *I* of $2_{ph}$ mode for **a-b,** 2H-TaS$_2$, and **c-d,** 2H-NbSe$_2$ respectively.

The β and *I* of two-phonon ($2_{ph}$) mode for both 2H-TaS$_2$ (Figure. S4a-b) and 2H-NbSe$_2$ (Figure. S4c-d) has a signature for its association with the CDW. All the mentioned data have abnormality below ~ 100 K, which indicates that CDW gap can persist even at higher than the reported CDW transition temperature. The persisted gap has a strong influence on the $2_{ph}$ mode, which is accompanying with both Kohn anomaly and CDW.